\documentstyle[11pt,twoside]{article}

\normalbaselines
\font\tbf = cmbx12

\begin{document}

\title{Higher-order geodesic deviations \\
applied to the Kerr metric}
\author{\tbf R. Colistete Jr.$^1$\thanks{%
e-mail: {\tt coliste@ccr.jussieu.fr}}, C. Leygnac$^2$\thanks{%
e-mail: {\tt leygnac@lapp.in2p3.fr}} and R. Kerner$^1$\thanks{%
e-mail: {\tt rk@ccr.jussieu.fr}} \\
\\
\mbox{\small $^1$Univ. Pierre et Marie Curie, L.P.T.L., CNRS URA 7600}\\
\mbox{\small Tour 22, 4\`eme \'etage, Bo\^{\i}te 142, 4 place Jussieu,
      75005 Paris, France}\\
\mbox{\small $^2$LAPTH, CNRS UMR 5108}\\
\mbox{\small Chemin de Bellevue, B.P. 110, 74941 Annecy-Le-Vieux Cedex,
France}}
\date{\today}
\maketitle

\begin{abstract}
Starting with an exact and simple geodesic, we generate approximate
geodesics by summing up higher-order geodesic deviations within a General
Relativistic setting, without using Newtonian and post-Newtonian
approximations.

We apply this method to the problem of closed orbital motion of test
particles in the Kerr metric space-time. With a simple circular orbit in the
equatorial plane taken as the initial geodesic we obtain finite eccentricity
orbits in the form of Taylor series with the eccentricity playing the role
of small parameter.

The explicit expressions of these higher-order geodesic deviations are
derived using successive systems of linear equations with constant
coefficients, whose solutions are of harmonic oscillator type. This scheme
gives best results when applied to the orbits with low eccentricities, but
with arbitrary values of $(GM/Rc^2)$, smaller than $1/6$ in the
Schwarzschild limit.
\end{abstract}

\vspace{0.7cm} PACS number(s): 04.25.--g, 04.25.Nx

\section{Introduction}

\label{introduction} \setcounter{equation}{0}

In two recently published articles, \cite{KHC} and \cite{BHK}, the
approximate low-eccentricity relativistic trajectories of planets with small
mass $m$ (as compared to the central body's mass $M$) have been constructed
in Schwarzschild and Reissner-Nordstr{\o}m space-time metrics. In the latter
case, the motion of electrically charged particles have been investigated,
too.

The two-body problem in General Relativity has been the object of many
excellent studies; one of the first checks of this theory has been the very
precise value of the perihelion advance calculated by Einstein \cite{AE}\
for the planet Mercury. The calculus was based on the solution of the
geodesic equation in Schwarzschild's metric, using the first integrals; the
solution was obtained in the form of a quadrature, with the proper time $%
\tau $ expressed as a quasi-elliptic integral. Such an integral can not be
evaluated analytically; instead, Einstein has developed the integrand into a
power series with respect to the small parameter $GM/rc^{2}$, which led to
simple integrations that could be easily performed. The approximate formula
for the perihelion advance after one revolution is then 
\begin{equation}
\Delta \,\varphi =\frac{6\pi GM}{a(1-e^{2})}=\frac{6\pi GM}{a}%
\,(1+e^{2}+e^{4}+e^{6}+...)  \label{peradvance1}
\end{equation}
where $a$ is the major semi-axis and $e$ the eccentricity of the orbit, and $%
G$ stands for Newton's gravitational constant divided by $c^{2}$, so that in
this notation the quantity $GM/r$ becomes dimensionless. We shall adopt this
notation from now on, equivalent of using the units in which $c=1$.\ The
above approximation is acceptable only if the value of the parameter $GM/a$
is negligibly small.

In our previous articles \cite{BHK} and \cite{KHC} we have proposed an
alternative way of determining the value of perihelion advance and finding
an explicit (parametrized by the proper time) form of trajectory and the law
of motion as a series of successive approximations, without supposing that
the ratio $GM/a$ (or equivalently, the ratio $v^{2}/c^{2}$) is very small,
or using the Newtonian approximation. Instead, we start from a very simple
particular solution of geodesic equation in Schwarzschild (or
Reissner-Nordstr{\o }m) metric: a perfect circular orbit along which the
small mass $m$ is advancing with a constant angular velocity. It is very
easy to check that such a motion is a geodesic curve in the aforementioned
space-times. The fact that all geometrical quantities, such as the
Christoffel coefficients, or the components of Riemann's tensor, take on 
{\it constant values} on this trajectory, leads to a particularly simple
form of the geodesic deviation equations: they reduce themselves to a system
of second-order linear differential equations with constant coefficients,
and the solution is just a collection of harmonic oscillators.

Here we apply the same method to the case of the axially-symmetric Kerr
metric. Although the motion of test particles along closed orbits in Kerr's
metric has been analyzed in a very exhaustive manner in many papers \cite
{Leaute}-\cite{Chandra}, our method gives results in an explicit form, and
is very well adapted for computer-based calculations. In the case of orbits
with low eccentricity it converges very quickly even for the non-negligible
values of the ratio $GM/r$.

In the following Sections we shall briefly recall the essential features of
our approximation method best suited for higher-order deviations. Using the
Kerr space-time, we choose a circular orbit in the equatorial plane as the
initial geodesic. Then, the first, second and third deviations are obtained,
the latter one with the help of the Poincar\'{e}'s method \cite{Poincare}
enabling us to obtain higher-order corrections to the basic frequencies. The
explicit form of the perihelion advance in the field of Kerr metric displays
interesting features as a combined result of the influence of two essential
parameters, the mass $M$ and the angular momentum density $a$ of the central
body.

In the last section, we discuss the physical content of the results and
consider some future applications of higher-order deviations, including the
effects of finite mass $m$ and internal spin (angular momentum) of the
planet. In contrast with our previous article \cite{KHC} we shall not
consider here the problem of gravitational radiation; it will be left to a
detailed future work.

\section{Geodesic deviations using small deformations}

The previous article \cite{KHC} was based on the deviation vectors $n^{\mu }$%
, $b^{\mu }$, $h^{\mu }$ and their deviation equations. However, as the
order of the deviation increases, it becomes harder to calculate the
deviation equations for the deviation vectors.

So, for our purpose here, which is the effective calculation of deformations
of circular orbits in Kerr metric, we need the explicit coordinate-dependent
expressions for the deviations that we shall add to given functions of
proper time $s$ which define the relativistic trajectory and the law of
motion. This is why we need to consider an alternative approach, which deals
with small deviations of arbitrary order of coordinate functions, thus
deforming the trajectories directly.

Consider an infinitesimal deformation of the geodesic curve $x^{\mu }\,(s)$: 
\begin{equation}
x^{\mu }\,(s)\Rightarrow {\tilde{x}}^{\mu }\,(s)=x^{\mu }\,(s)+\delta x^{\mu
}\,(s).  \label{tilde1}
\end{equation}
Suppose that we want the new curve ${\tilde{x}}^{\mu }(s)$ to satisfy the
geodesic equation, too: 
\begin{equation}
\frac{d^{2}{\tilde{x}}^{\mu }}{ds^{2}}+\Gamma _{\lambda \rho }^{\mu }({%
\tilde{x}}^{\nu })\,\frac{d{\tilde{x}}^{\lambda }}{ds}\,\frac{{d\tilde{x}}%
^{\rho }}{ds}=0.  \label{geodevtilde}
\end{equation}
Expanding the Christoffel coefficients into power series of $\delta x^{\mu }$%
, 
\begin{equation}
\Gamma _{\lambda \rho }^{\mu }({\tilde{x}}^{\nu })=\Gamma _{\lambda \rho
}^{\mu }(x^{\nu })+\delta x^{\sigma }\,\partial _{\sigma }\,\Gamma _{\lambda
\rho }^{\mu }(x^{\nu })+\frac{1}{2!}\,\delta x^{\sigma }\,\delta x^{\tau
}\,\partial _{\sigma \tau }^{2}\,\Gamma _{\lambda \rho }^{\mu }(x^{\nu
})+\dots  \label{Taylor1}
\end{equation}
Substituting the expressions (\ref{tilde1}) and (\ref{Taylor1}) into Eq. (%
\ref{geodevtilde}) and expanding in consecutive powers of de\-via\-tions $%
\delta x^{\mu }$, we get in the zeroth-order the initial geodesic equation (%
\ref{geodevtilde}) satisfied by $x^{\mu }(s)$; collecting then all the terms
linear in $\delta x^{\mu }$ and their derivatives, we get 
\begin{equation}
\frac{d^{2}\delta x^{\mu }}{ds^{2}}+2\,\Gamma _{\lambda \rho }^{\mu
}\,u^{\lambda }\,\frac{d\delta x^{\rho }}{ds}+(\partial _{\sigma }\,\Gamma
_{\lambda \rho }^{\mu })\,u^{\lambda }u^{\rho }\delta x^{\sigma }=0
\label{deltageodev1}
\end{equation}
which coincides with the first-order deviation equation of the Ref.~\cite
{KHC} if we replace the vector $n^{\mu }$ by the infinitesimal deviation $%
\delta x^{\mu }$. The geometrical meaning of this equation is now very
clear: it gives the conditions to be satisfied by infinitesimal functions $%
\epsilon \,n^{\mu }\,(s)=\delta x^{\mu }$ (with $\epsilon $ being an
infinitesimal parameter) defined along a given geodesic curve $x^{\lambda
}\,(s)$, in order to ensure that the new curve, infinitesimally close to it
and defined by ${\tilde{x}}^{\lambda }\,(s)=x^{\lambda }\,(s)+\epsilon
\,n^{\lambda }\,(s)$ , is also a geodesic one, up to the first order in $%
\epsilon $.

The fact that $n^{\mu }$ is a vector is in agreement with the transformation
properties of {\it infinitesimal} deviations $\delta x^{\mu }$, which under
an arbitrary change of coordinates $x^{\mu }=x^{\mu }\,(y^{\rho })$
transform as 
\begin{equation}
\delta x^{\mu }=\frac{\partial x^{\mu }}{\partial y^{\lambda }}\,\delta
y^{\lambda }
\end{equation}
up to higher-order terms, neglected at the linear approximation level.

However, the higher-order terms in the expansion are still there: collecting
all the second-order terms in $\delta x^{\mu }$\ from the Taylor expansion
of Eq. (\ref{geodevtilde}), we get 
\begin{equation}
\Gamma _{\lambda \rho }^{\mu }\,\frac{d\delta x^{\lambda }}{ds}\,\frac{%
d\delta x^{\rho }}{ds}+2\,\delta x^{\nu }\,(\partial _{\nu }\,\Gamma
_{\lambda \rho }^{\mu })\,u^{\lambda }\,\frac{d\delta x^{\rho }}{ds}+\frac{1%
}{2}\,\delta x^{\nu }\,\delta x^{\sigma }\,(\partial _{\nu }\partial
_{\sigma }\,\Gamma _{\lambda \rho }^{\mu })\,u^{\lambda }u^{\rho },
\label{Taylor2}
\end{equation}
and there is no reason for it to vanish even if the deviations $\delta
x^{\mu }(s)$ satisfy Eq. (\ref{deltageodev1}). The vanishing of expression (%
\ref{Taylor2}) would impose too many conditions, {\it a priori} incompatible
with Eq. (\ref{deltageodev1}) on the same set of functions $\delta x^{\mu
}(s)$, and we need extra degrees of freedom if we want to cancel also all
the second-order terms.

This means that from the very beginning, infinitesimal deviations of
higher-order must be introduced: 
\begin{equation}
{\tilde{x}}^{\mu }\,(s)=x^{\mu }\,(s)+\delta x^{\mu }\,(s)+\frac{1}{2!}%
\,\delta ^{2}x^{\mu }\,(s)+\frac{1}{3!}\,\delta ^{3}x^{\mu }\,(s)+\dots
\label{tilde2}
\end{equation}
so that two new second-order terms will add up to the expression (\ref
{Taylor2}), namely 
\begin{equation}
\frac{d^{2}(\delta ^{2}x^{\mu })}{ds^{2}}+\delta ^{2}x^{\nu }\,\partial
_{\nu }\,\Gamma _{\lambda \rho }^{\mu }\,u^{\lambda }u^{\rho }
\end{equation}
and the sum of all these terms represents a new set of second-order
differential equations imposed on the independent functions $\delta
^{2}x^{\mu }$, which may be solved after we insert the solutions for $u^{\mu
}(s)$ and $\delta x^{\mu }(s)$ obtained previously.

At this point, another problem arises: not only the coefficients in these
differential equations are not covariant objects, but also the functions $%
\delta ^{2}x^{\mu }$ do not behave as vectors under the change of
coordinates. As a matter of fact, they will mix up with terms quadratic in $%
\delta x^{\mu }$ as follows: 
\begin{equation}
\delta ^{2}x^{\mu }=\delta (\delta x^{\mu })=\frac{\partial x^{\mu }}{%
\partial y^{\lambda }}\,\delta ^{2}y^{\lambda }+\frac{\partial ^{2}x^{\mu }}{%
\partial y^{\lambda }\,\partial y^{\rho }}\,\delta y^{\lambda }\,\delta
y^{\rho }\,.  \label{transform2}
\end{equation}
This non-homogeneous transformation law suggests that we can introduce a
covariant quantity $D^{2}x^{\mu }$ defined as 
\begin{equation}
D^{2}x^{\mu }=\delta ^{2}x^{\mu }+\Gamma _{\lambda \rho }^{\mu }\,\delta
x^{\lambda }\,\delta x^{\rho }.  \label{covdelta2}
\end{equation}
Defining the infinitesimal vector $b^{\mu }$ as $\epsilon ^{2}\,b^{\mu
}=D^{2}x^{\mu }$ and expressing the Taylor expansion (\ref{tilde2}) in terms
of $n^{\mu }$ and $b^{\mu }$ as 
\begin{equation}
{\tilde{x}}^{\mu }=x^{\mu }+\epsilon \,n^{\mu }+\frac{1}{2!}\epsilon
^{2}\left( b^{\mu }-\Gamma _{\lambda \rho }^{\mu }\,n^{\lambda }n^{\rho
}\right) +\dots  \label{covtilde}
\end{equation}
and requiring the geodesic equation for ${\tilde{x}}^{\mu }$ to be satisfied
up to the second order in $\epsilon $, we arrive at the same second-order
deviation equation of the Ref. \cite{KHC} satisfied by $b^{\mu }$ that is
non-manifestly covariant and equivalent to the manifestly covariant equation.

In Ref.~\cite{KHC}, the non-manifestly covariant deviation equations could
be easier obtained using this Taylor expansion approach than transforming
the manifestly covariant deviation equations. In practical calculations the
non-manifestly covariant equations turn out to be of much use, i.e., it is
easier to obtain the solutions of $n^{\mu }$ and $b^{\mu }$ using them
instead of the manifestly covariant equations.

Similar corrections are needed to define the higher-order deviations, like 
\begin{equation}
\epsilon ^{3}h^{\mu }=D^{3}x^{\mu }=\delta ^{3}x^{\mu }+3\Gamma _{\nu \sigma
}^{\mu }\,\delta x^{\nu }\,\delta ^{2}x^{\sigma }+\left( \partial _{\lambda
}\,\Gamma _{\nu \sigma }^{\mu }+\Gamma _{\lambda \rho }^{\mu }\,\Gamma _{\nu
\sigma }^{\rho }\right) \delta x^{\lambda }\,\delta x^{\nu }\,\delta
x^{\sigma },  \label{covdelta3}
\end{equation}
so the real coordinate deviation $\delta ^{3}x^{\mu }$ reads 
\begin{equation}
\delta ^{3}x^{\mu }=\epsilon ^{3}[h^{\mu }-3\Gamma _{\rho \sigma }^{\mu
}\,n^{\rho }b^{\sigma }-\left( \partial _{\lambda }\,\Gamma _{\nu \sigma
}^{\mu }-2\Gamma _{\lambda \rho }^{\mu }\,\Gamma _{\nu \sigma }^{\rho
}\right) n^{\lambda }n^{\nu }n^{\sigma }].  \label{delta3}
\end{equation}
A study of higher-order differentials and their covariant generalizations
can be found in recent papers \cite{Kerner98}-\cite{Abramov}.

But there is even another easier form to calculate higher-order geodesic
deviations : we indeed need $\delta ^{n}x^{\mu }$ to obtain the geodesic ${%
\tilde{x}}^{\mu }$, and the differential equations for $\delta ^{n}x^{\mu }$
are simpler than their counterparts $n^{\mu }$, $b^{\mu }$, $h^{\mu }$, etc.
For example, requiring again the geodesic equation for ${\tilde{x}}^{\mu }$
to be satisfied up to the second order in $\epsilon $, the following
second-order deviation equation for $\delta ^{2}x^{\mu }$ is obtained 
\[
{\frac{d^{2}\delta ^{2}x^{\mu }}{ds^{2}}+(\partial _{\rho }\,\Gamma
_{\lambda \sigma }^{\mu })\,u^{\lambda }u^{\sigma }\delta ^{2}x^{\rho
}+2\,\Gamma _{\lambda \sigma }^{\mu }\,u^{\lambda }\,\frac{d\delta
^{2}x^{\sigma }}{ds}=}
\]
\begin{equation}
{-2\,\Gamma _{\lambda \rho }^{\mu }\frac{d\delta x^{\lambda }}{ds}\frac{%
d\delta x^{\rho }}{ds}-4(\partial _{\sigma }\,\Gamma _{\lambda \rho }^{\mu
})\,u^{\lambda }\,\delta x^{\sigma }\frac{d\delta x^{\rho }}{ds}-(\partial
_{\nu }\,\partial _{\sigma }\,\Gamma _{\lambda \rho }^{\mu })\,u^{\lambda
}\,u^{\rho }\,\delta x^{\sigma }\,\delta x^{\nu },}  \label{deltageodev2}
\end{equation}
where we see that the l.h.s. is unchanged, but the r.h.s. has only $3$ terms
instead of $10$ found in the non-manifestly covariant second-order deviation
equation for $b^{\mu }$ (see Ref. \cite{KHC}).

The non-manifestly covariant third-order deviation equation for $h^{\mu }$
is not shown here, but has $60$ terms in the r.h.s., while the third-order
deviation equation for $\delta ^{3}x^{\mu }$ has only $7$ terms: 
\[
{\frac{d^{2}\delta ^{3}x^{\mu }}{ds^{2}}+(\partial _{\rho }\,\Gamma
_{\lambda \sigma }^{\mu })\,u^{\lambda }u^{\sigma }\delta ^{3}x^{\rho
}+2\,\Gamma _{\lambda \sigma }^{\mu }\,u^{\lambda }\,\frac{d\delta
^{3}x^{\sigma }}{ds}=} 
\]
\[
{-6\,\Gamma _{\lambda \rho }^{\mu }\frac{d\delta ^{2}x^{\lambda }}{ds}\frac{%
d\delta x^{\rho }}{ds}-6(\partial _{\sigma }\,\Gamma _{\lambda \rho }^{\mu })%
}\left( {\delta x^{\sigma }\frac{d\delta x^{\lambda }}{ds}\frac{d\delta
x^{\rho }}{ds}}+{u^{\lambda }\,\delta }^{2}{x^{\sigma }\frac{d\delta x^{\rho
}}{ds}+u^{\lambda }\,\delta x^{\sigma }\frac{d\delta ^{2}x^{\rho }}{ds}}%
\right) 
\]
\begin{equation}
{-3(\partial _{\nu }\,\partial _{\sigma }\,\Gamma _{\lambda \rho }^{\mu
})\,u^{\lambda }\,\delta x^{\nu }}\left( {2\,\delta x^{\sigma }\frac{d\delta
x^{\rho }}{ds}+u^{\rho }\,}\delta ^{2}x^{\sigma }\right) {-(\partial _{\tau
}\,\partial _{\nu }\,\partial _{\sigma }\,\Gamma _{\lambda \rho }^{\mu
})\,u^{\lambda }\,u^{\rho }\,\delta x^{\sigma }\,\delta x^{\nu }\,\delta
x^{\tau }.}  \label{deltageodev3}
\end{equation}

The fourth-order deviation equation for $\delta ^{4}x^{\mu }$ has $15$
terms, and the fifth-order deviation equation for $\delta ^{5}x^{\mu }$ has $%
26$ terms. We have developed a symbolic computer program to calculate $%
n^{th} $-order deviation equations for $\delta ^{n}x^{\mu }$.

The non-manifestly covariant geodesic deviation\ equations are well suited
to deriving successive approximations for geodesics close to an initial one.
Starting from a given geodesic $x^{\mu }\,(s)$ we can solve Eq. (\ref
{deltageodev1}) and find the first-order deviation vector $\delta x^{\mu
}\,(s)$. Now, with $u^{\mu }\,(s)$ and $\delta x^{\mu }\,(s)$, the system (%
\ref{deltageodev2}) can be solved and we obtain the second-order deviation $%
\delta ^{2}x^{\mu }\,(s)$.Then, using $u^{\mu }\,(s)$, $\delta x^{\mu }\,(s)$
and $\delta ^{2}x^{\mu }\,(s)$ into the system (\ref{deltageodev3}) the
third-order deviation $\delta ^{3}x^{\mu }\,(s)$ is calculated, and so forth.

The literature about geodesic deviations includes a rigorous mathematical
study of geodesic deviations up to the second-order, as well as geometric
interpretation, but using different derivation, presented in \cite{Bazanski1}%
. Also, a Hamilton--Jacobi formalism had been derived in \cite{Bazanski2},
which was applied to the problem of free falling particles in the
Schwarzschild space-time \cite{BazansJara}. Anyway, the resulting
expressions are not well optimized for successive calculations of
higher-order geodesic deviations. Interesting effects resulting from the
analysis of first-order geodesic deviations of test particles suspended in
hollow spherical satellites have been discussed in \cite{Shirokov}.

\section{Circular orbits in the Kerr metric}

We choose as initial geodesic a circular orbit in the axisymmetric
gravitational field created by a massive body with rotation, i.e., in the
Kerr metric. This metric and their circular orbits have been studied in
several papers \cite{Leaute}-\cite{Wilkins} and books \cite{MTW}-\cite
{Chandra}.

The gravitational field is described by the line-element (in natural
coordinates with $c=1$ and $G=1$) 
\begin{equation}
ds^{2}=\frac{\rho ^{2}}{\Delta }\,dr^{2}+\rho ^{2}d\theta ^{2}+\frac{%
2Mr-\rho ^{2}}{\rho ^{2}}\,dt^{2}-\frac{4Mra}{\rho ^{2}}\,{sin}^{2}\theta
\,dt\,d\phi +\frac{{sin}^{2}\theta }{\rho ^{2}}\,(-\Delta \,a^{2}\,{sin}%
^{2}\theta +(r^{2}+a^{2})^{2})\,d\phi ^{2}  \nonumber
\end{equation}
with $\Delta =r^{2}+a^{2}-2\,M\,r$ and $\rho ^{2}=r^{2}+a^{2}\,{\sin }%
^{2}\theta $, where $M$ and $a=\frac{J}{M}$ are the mass and the angular
momentum density of the central body rotating in opposite $\phi $ direction.

The circular orbit of radius $R$ in the equatorial plane (which is a
geodesic in the background Kerr metric) is described by a simple $4$%
-velocity vector: 
\[
u^{r}=\frac{dr}{ds}=0,\,\,\ \ u^{\theta }=\frac{d\theta }{ds}=0, 
\]
\[
u^{\phi }=\frac{d\phi }{ds}=\omega _{c}=\frac{\frac{\sqrt{M}}{R^{3/2}}}{%
\sqrt{1-\frac{3M}{R}+\frac{2a\sqrt{M}}{R^{3/2}}\,}}, 
\]
\begin{equation}
u^{t}=\frac{dt}{ds}=\frac{1+\frac{a\sqrt{M}}{R^{3/2}}}{\sqrt{1-\frac{3M}{R}+%
\frac{2a\sqrt{M}}{R^{3/2}}}},  \label{4velocity}
\end{equation}
because $r=R=\mbox{const.}$, $\ \ \theta =\pi /2=\mbox{const.}$, so that $%
\sin \,\theta =1$ \ \ and $\ \ \cos \,\theta =0$. The angular frequency of
this circular motion is $\omega _{c}/u^{t}$, which decreases if the central
body increases the rotation in the opposite direction.

\section{First-order geodesic deviation around a circular orbit}

Likewise the calculations for the Schwarzschild case in Ref.~\cite{KHC}, we
use the non-manifestly first-order deviation equation for $\delta x^{\mu }$,
Eq. (\ref{deltageodev1}). The vector $u^{\mu }$ of a circular orbit,
calculated before, is used, providing constant components that yield simple
equations for the components $\delta r$, $\delta \theta $, $\delta \phi $
and $\delta t$ (or $n^{r}$, $n^{\theta }$, $n^{\phi }$ and $n^{t}$), as we
can see in a matrix form: 
\begin{equation}
\pmatrix{m_{11} & m_{12} & m_{13} & m_{14}\cr m_{21} & m_{22} & m_{23} &
m_{24} \cr m_{31} & m_{32} & m_{33} & m_{34} \cr m_{41} & m_{42} & m_{43} &
m_{44}}\,\,\pmatrix{\delta r \cr \delta \theta \cr \delta \phi \cr \delta t}%
\,=\,\pmatrix{0 \cr 0 \cr 0 \cr 0},  \label{matrixdeltax}
\end{equation}
where the matrix elements are: 
\begin{equation}
m_{11}=\frac{d^{2}}{ds^{2}}-\frac{3M}{R^{3}}\frac{f_{2}}{f_{1}},\,\,\ \
m_{12}=0,\,\,\ \ m_{13}=-2\sqrt{\frac{M}{R}}\left( 1+\frac{a\sqrt{M}}{R^{3/2}%
}\right) \frac{f_{2}}{\sqrt{f_{1}}}\frac{d}{ds},\,\,\ \ m_{14}=\frac{2M}{%
R^{2}}\frac{f_{2}}{\sqrt{f_{1}}}\frac{d}{ds},  \label{m1j}
\end{equation}
\begin{equation}
m_{21}=m_{23}=m_{24}=0,\,\,\ \ m_{22}=\frac{d^{2}}{ds^{2}}+\frac{M}{R^{3}}%
\frac{f_{3}}{f_{1}},  \label{m2j}
\end{equation}
\begin{equation}
m_{31}=\frac{2\sqrt{M}}{R^{5/2}}\frac{f_{4}}{f_{2}\sqrt{f_{1}}}\frac{d}{ds}%
,\,\,\ \ m_{32}=m_{34}=0,\,\,\ \ m_{33}=\frac{d^{2}}{ds^{2}},  \label{m3j}
\end{equation}
\begin{equation}
m_{41}=\frac{2M}{R^{2}}\frac{f_{5}}{f_{2}\sqrt{f_{1}}}\frac{d}{ds},\,\,\ \
m_{42}=m_{43}=0,\,\,\ \ m_{44}=\frac{d^{2}}{ds^{2}},  \label{m4j}
\end{equation}
using the functions: 
\[
f_{1}=\left( 1-\frac{3M}{R}\right) +\frac{2a\sqrt{M}}{R^{3/2}},\,\,\ \
f_{2}=\left( 1-\frac{2M}{R}\right) +\frac{a^{2}}{R^{2}},\,\,\ \ f_{3}=1-%
\frac{4a\sqrt{M}}{R^{3/2}}+\frac{3a^{2}}{R^{2}}, 
\]
\begin{equation}
f_{4}=\left( 1-\frac{2M}{R}\right) +\frac{a\sqrt{M}}{R^{3/2}},\,\,\ \
f_{5}=1-\frac{2a\sqrt{M}}{R^{3/2}}+\frac{a^{2}}{R^{2}}.  \label{f1f5}
\end{equation}

The harmonic oscillator equation for $n^{\theta }=\delta \theta $ has an
angular frequency $\omega _{\theta }$:

\begin{equation}
\omega _{\theta }=\frac{\sqrt{M}}{R^{3/2}}\sqrt{\frac{f_{3}}{f_{1}}}=\frac{%
\sqrt{M}}{R^{3/2}}\sqrt{\frac{1-\frac{4a\sqrt{M}}{R^{3/2}}+\frac{3a^{2}}{%
R^{2}}}{\left( 1-\frac{3M}{R}\right) +\frac{2a\sqrt{M}}{R^{3/2}}}}.
\label{omegatetha}
\end{equation}
One possible choice of solution is 
\begin{equation}
n^{\theta }=\delta \theta =-n_{0}^{\theta }\cos (\omega _{\theta }s).
\label{ntheta}
\end{equation}
In the Schwarzschild limit ($a\rightarrow 0$), $\omega _{\theta }=\omega
_{c} $, so in this case we can neglect this solution ($n_{0}^{\theta }=0$)
because the new plane of orbit is a new one inclined, or just a change of
coordinate system.

Using the differential equation for $n^{r}=\delta r$ we can eliminate the
derivatives of $\delta \phi $ and $\delta t$, yielding the harmonic
oscillator equation 
\begin{equation}
\frac{d^{2}\delta r}{ds^{2}}+\omega ^{2}\delta r=0,  \label{hoedeltar}
\end{equation}
with the characteristic frequency 
\begin{equation}
\omega =\frac{\sqrt{M}}{R^{3/2}}\sqrt{\frac{\left( 1-\frac{6M}{R}\right) +%
\frac{8a\sqrt{M}}{R^{3/2}}-\frac{3a^{2}}{R^{2}}}{\left( 1-\frac{3M}{R}%
\right) +\frac{2a\sqrt{M}}{R^{3/2}}}}=\frac{\sqrt{M}}{R^{3/2}}\sqrt{\frac{%
f_{6}}{f_{1}}},  \label{omega}
\end{equation}
where: 
\begin{equation}
f_{6}=\left( 1-\frac{6M}{R}\right) +\frac{8a\sqrt{M}}{R^{3/2}}-\frac{3a^{2}}{%
R^{2}}.  \label{f6}
\end{equation}

We shall choose the initial phase to have (with $n_{0}^{r}>0$): 
\begin{equation}
n^{r}=\delta r=-n_{0}^{r}\cos (\omega s)  \label{nr}
\end{equation}
so the perihelion occurs when $s=0$.

The calculation of $\delta \phi $ and $\delta t$ is now simple: 
\begin{equation}
n^{\phi }=\delta \phi =n_{0}^{\phi }\sin (\omega s),  \label{nphi}
\end{equation}
\begin{equation}
n^{t}=\delta t=n_{0}^{t}\sin (\omega s),  \label{nt}
\end{equation}
where the amplitudes depend on $n_{0}^{r}$: 
\begin{equation}
n_{0}^{\phi }=\frac{2n_{0}^{r}}{R}\frac{f_{4}}{f_{2}\sqrt{f_{6}}},
\label{nphi0}
\end{equation}
\begin{equation}
n_{0}^{t}=2n_{0}^{r}\sqrt{\frac{M}{R}}\frac{f_{5}}{f_{2}\sqrt{f_{6}}}.
\label{nt0}
\end{equation}

Adding this first-order deviation to the circular orbit, the new trajectory
and the law of motion are given by 
\begin{eqnarray}
r &=&R-\,n_{0}^{r}\,\cos (\omega s),  \label{r1order} \\
\theta &=&\frac{\pi }{2}-\,n_{0}^{\theta }\,\cos (\omega _{\theta }s),
\label{theta1order} \\
\varphi &=&\omega _{c}\,s+\,n_{0}^{\varphi }\,\sin (\omega \,s),
\label{phi1order} \\
t &=&u^{t}s+\,n_{0}^{t}\,\sin (\omega \,s),  \label{t1order}
\end{eqnarray}
and this solution is a geodesic up to the first-order in $\epsilon $. It is
important to note once again that the coefficient $n_{0}^{r}$, which also
fixes the values of the two remaining amplitudes, $n_{0}^{t}$ and $%
n_{0}^{\varphi }$, defines the size of the actual deviation, so that the
ratio $\frac{n_{0}^{r}}{R}$ becomes the dimensionless infinitesimal
parameter $\epsilon $ controlling the approximation series with consecutive
terms proportional to the consecutive powers of $\frac{n_{0}^{r}}{R}$ (or$%
\frac{n_{0}^{\theta }}{R}$).

One easily checks that the Schwarzschild limit ($a\rightarrow 0$) of the
solution above yields the results of the Ref. \cite{KHC}, including the
perihelion advance and the generalized epicycle \cite{Ptolemaios}, where we
identify the major semi-axis $a$ with $R$ and the\ eccentricity $e$ with $%
\frac{n_{0}^{r}}{R}$ : 
\begin{equation}
r(t)=\frac{a(1-e^{2})}{1+e\,\cos (\omega _{0}\,t)}\simeq a\,\left[ 1-e\,\cos
(\omega _{0}\,t)\right] ,  \label{ellipse}
\end{equation}
But it is more interesting to show the perihelion advance for the Kerr case,
even using the first-order deviation, 
\begin{eqnarray}
\Delta \varphi &=&2\,\pi \left( \frac{\omega _{c}}{\omega }-1\right) =\left( 
\frac{6\,\pi M}{R}+\frac{27\,\pi M^{2}}{R^{2}}+\frac{135\,\pi M^{3}}{R^{3}}%
+...\right) -a\left( \frac{8\,\pi \sqrt{M}}{R^{3/2}}+\frac{72\,\pi M^{3/2}}{%
R^{5/2}}\right.  \nonumber \\
&&\left. +\frac{540\,\pi M^{5/2}}{R^{7/2}}+...\right) +a^{2}\left( \frac{%
3\,\pi }{R^{2}}+\frac{75\,\pi M}{R^{3}}+\frac{1845\,\pi M^{2}}{2R^{4}}%
+...\right) +...  \label{deltaphi0}
\end{eqnarray}
because the Kerr parameter $a$ appears with positive and negative
coefficients, i.e., the angular momentum density $a$ can increase or
decrease the perihelion advance. Note that the post-Newtonian limit matches
the Eq. (\ref{peradvance1}) for small eccentricities.

Despite the limitations of the first-order geodesic deviation, we have
already obtained a ge\-ne\-ra\-li\-zed perihelion advance valid for high
values of $\frac{M}{R}$ and $a$, but low values of the ``eccentricity'' $e$
(or $\frac{n_{0}^{r}}{R}$). The high-order deviations will, for example,
allow the calculation of $\Delta \varphi $ for higher values of the
``eccentricity'' $\frac{n_{0}^{r}}{R}$.

\section{The second-order geodesic deviation}

Inserting the complete solution for the first-order deviation $\delta x^{\mu
}=n^{\mu }$, Eqs. (\ref{ntheta}), (\ref{nr})--(\ref{nt}) into the
second-order deviation equation for $\delta ^{2}x^{\mu }$ (\ref{deltageodev2}%
), we find the same matrix with a new non-homogeneous vector $C^{\mu }$: 
\begin{equation}
\pmatrix{ m_{11} & m_{12} & m_{13} & m_{14}\cr m_{21} & m_{22} & m_{23} &
m_{24} \cr m_{31} & m_{32} & m_{33} & m_{34} \cr m_{41} & m_{42} & m_{43} &
m_{44}}\,\,\pmatrix{\delta ^{2} r \cr \delta ^{2} \theta \cr \delta ^{2}
\phi \cr \delta ^{2} t}=\epsilon ^{2}\,\pmatrix{C^r \cr C^\theta \cr C^\phi
\cr C^t}\,,  \label{matrixdelta2x}
\end{equation}
where we have put into evidence the common factor $\epsilon ^{2}$, which
shows the explicit quadratic dependence of the second-order deviation $%
\delta ^{2}x^{\mu }$ on the first-order deviation amplitude $n_{0}^{r}$ (or $%
n_{0}^{\theta }$). The functions $C^{r},C^{\theta }$, $C^{\phi }$ and $C^{t}$
are expressions depending on $M$, $R$, $a$, and on the functions $\sin
(2\omega s)$, $\cos (2\omega s)$, $\sin (2\omega _{\theta }s)$, $\cos
(2\omega _{\theta }s)$, $\cos \left[ (\omega -\omega _{\theta })s\right] $
and $\cos \left[ (\omega +\omega _{\theta })s\right] $: 
\begin{equation}
C^{r}=C_{0}^{r}+C_{2r}^{r}\cos (2\omega s)+C_{2\theta }^{r}\cos (2\omega
_{\theta }s),  \label{Cr}
\end{equation}
\begin{equation}
C^{\theta }=C_{-}^{\theta }\cos [(\omega -\omega _{\theta })s]+C_{+}^{\theta
}\cos [(\omega +\omega _{\theta })s],  \label{Ctheta}
\end{equation}
\begin{equation}
C^{\phi }=C_{2r}^{\phi }\sin (2\omega s)+C_{2\theta }^{\phi }\sin (2\omega
_{\theta }s),  \label{Cphi}
\end{equation}
\begin{equation}
C^{t}=C_{2r}^{t}\sin (2\omega s)+C_{2\theta }^{t}\sin (2\omega _{\theta }s).
\label{Ct}
\end{equation}

The solution of the above matrix for $\delta ^{2}x^{\mu }(s)$ has the same
characteristic equations of the matrix (\ref{matrixdeltax}) for $\delta
x^{\mu }(s)=n^{\mu }(s)$, and the general solution containing oscillating
terms with\ angular frequency $\omega $ and $\omega _{\theta }$\ is of no
interest because it is already accounted for by $n^{\mu }(s)$. But the
particular solution includes the terms linear in the proper time $s$,
constant ones, and the terms oscillating with\ angular frequency $2\omega $, 
$(\omega -\omega _{\theta })$ and $(\omega +\omega _{\theta })$:

\begin{equation}
\delta ^{2}r=\delta ^{2}r_{0}+\delta ^{2}r_{2}\cos (2\omega s),  \label{d2r}
\end{equation}
\begin{equation}
\delta ^{2}\theta =\delta ^{2}\theta _{-}\cos [(\omega -\omega _{\theta
})s]+\delta ^{2}\theta _{+}\cos [(\omega +\omega _{\theta })s],
\label{d2theta}
\end{equation}
\begin{equation}
\delta ^{2}\phi =(\delta ^{2}\phi _{0})s+\delta ^{2}\phi _{2r}\sin (2\omega
s)+\delta ^{2}\phi _{2\theta }\sin [2\omega _{\theta }s],  \label{d2phi}
\end{equation}
\begin{equation}
\delta ^{2}t=(\delta ^{2}t_{0})s+\delta ^{2}t_{2r}\sin (2\omega s)+\delta
^{2}t_{2\theta }\sin [2\omega _{\theta }s].  \label{d2t}
\end{equation}
The constants $\delta ^{2}r_{0}$, $\delta ^{2}\phi _{0}$ and $\delta
^{2}t_{0}$ depend on two arbitrary constants, so we can choose the initial
conditions of the differential solutions so that the constants $\delta
^{2}r_{0}$ and $\delta ^{2}\phi _{0}$ are null, and $\delta ^{2}t_{0}$ is
simplified. The Appendix 1 shows the explicit values of the above
coefficients.

In the Schwarzschild limit, the solution for the second-order geodesic
deviation $\delta ^{2}x^{\mu }(s)$ is: 
\begin{equation}
\delta ^{2}r=-\frac{\left( n_{0}^{r}\right) ^{2}}{R}\frac{\left( 1-\frac{7M}{%
R}\right) }{\left( 1-\frac{6M}{R}\right) }\cos (2\omega s),\,\,\ \ \delta
^{2}\theta =0,\,\,\ \ \delta ^{2}\phi =-2\frac{\left( n_{0}^{r}\right) ^{2}}{%
R^{2}}\frac{\left( 5-\frac{32M}{R}\right) }{\left( 1-\frac{6M}{R}\right)
^{3/2}}\sin (2\omega s),  \label{d2rd2thetad2varphiS}
\end{equation}
\begin{equation}
\delta ^{2}t=\frac{\left( n_{0}^{r}\right) ^{2}}{R}\left[ -\frac{3}{2}\frac{%
\left( 1+\frac{M}{R}\right) }{R\left( 1-\frac{2M}{R}\right) \sqrt{1-\frac{3M%
}{R}}}s+\sqrt{\frac{M}{R}}\frac{2-\frac{15M}{R}+\frac{14M^{2}}{R^{2}}}{(1-%
\frac{2M}{R})^{2}\left( 1-\frac{6M}{R}\right) ^{3/2}}\sin (2\omega s)\right]
.  \label{d2tS}
\end{equation}
The second-order deviation $\delta ^{2}x^{\mu }$ computed in the Ref. \cite
{KHC} used another choice for the constants $\delta ^{2}r_{0}$, $\delta
^{2}\phi _{0}$ and $\delta ^{2}t_{0}$, i.e., equivalent to different initial
conditions for the initial geodesic. So, using the initial conditions chosen
above, the comparison with an ellipse in the Schwarzschild limit also gives
more compact results.

The trajectory described by $x^{\mu }$ including second-order deviations is
not an ellipse due to the General Relativity effects of $\frac{M}{R}$, but
we can match the perihelion and aphelion distances of the Keplerian, i.e.,
elliptical orbit, with the same perihelion and aphelion distances of the
orbit described by $x^{\mu }$: 
\begin{equation}
a=R-\frac{\left( n_{0}^{r}\right) ^{2}}{2R}\frac{\left( 1-\frac{7M}{R}%
\right) }{\left( 1-\frac{6M}{R}\right) },  \label{a2order}
\end{equation}
\begin{equation}
e=\frac{2n_{0}^{r}\left( 1-\frac{6M}{R}\right) }{2R\left( 1-\frac{6M}{R}%
\right) -\frac{\left( n_{0}^{r}\right) ^{2}}{R}\left( 1-\frac{7M}{R}\right) }%
=\frac{n_{0}^{r}}{R}+{\cal O}\left( \frac{(n_{0}^{r})^{3}}{R^{3}}\right) .
\label{e2order}
\end{equation}
The shape of the orbit described by $r(\varphi )$ can be obtained from $%
\varphi (s)$, then $s(\varphi )$ by means of successive approximations
beginning with $\omega s=\frac{\omega }{\omega _{c}}\varphi $, and $s$ is
replaced in $r(s)$ giving $r(\varphi )$ up to the second order in $\frac{%
n_{0}^{r}}{R}$: 
\begin{equation}
\frac{r}{R}=1-\frac{n_{0}^{r}}{R}\cos \left( \frac{\omega }{\omega _{c}}%
\varphi \right) +\left( \frac{n_{0}^{r}}{R}\right) ^{2}\left[ -1+\frac{%
\left( 1-\frac{5M}{R}\right) }{2\left( 1-\frac{6M}{R}\right) }\cos \left( 
\frac{2\omega }{\omega _{c}}\varphi \right) \right] +...  \label{r2order}
\end{equation}
The exact equation of an ellipse is obtained in the limit $\frac{M}{R}%
\rightarrow 0$, up to the second order in $e=n_{0}^{r}/R$: 
\begin{equation}
r=\frac{r_{0}}{1+e\,\cos \varphi }=\frac{\left( 1-\frac{3}{2}e^{2}\right) R}{%
1+e\,\cos \varphi }=R\left[ 1-e\cos \varphi +e^{2}\left( -1+\frac{1}{2}\cos
2\varphi \right) +...\right] .  \label{ellipser0}
\end{equation}
In the ellipse equation (\ref{ellipse}) we have $r_{0}=a(1-e^{2})$, so 
\begin{equation}
a=\frac{R\left( 1-\frac{3}{2}e^{2}\right) }{\left( 1-e^{2}\right) }\simeq
R\left( 1-\frac{e^{2}}{2}\right) .  \label{semiaxisa}
\end{equation}
These values for $a$ and $e$ agree with Eqs. (\ref{a2order})--(\ref{e2order}%
).

In order to improve the comparison of the perihelion advance in the
post-Newtonian limit with Eq. (\ref{peradvance1}), the $\Delta \varphi $
should include $\frac{n_{0}^{r}}{R}$ terms, which is not yet the case using
second-order deviations due to the imposed initial conditions.

\section{Third-order deviation and Poincar\'{e}'s method}

Using the solutions for $\delta x^{\mu }=n^{\mu }$ and $\delta ^{2}x^{\mu }$
into third-order deviation equation for $\delta ^{3}x^{\mu }$ (\ref
{deltageodev3}), we again find the same matrix with a new non-homogeneous
vector $D^{\mu }$: 
\begin{equation}
\pmatrix{ m_{11} & m_{12} & m_{13} & m_{14}\cr m_{21} & m_{22} & m_{23} &
m_{24} \cr m_{31} & m_{32} & m_{33} & m_{34} \cr m_{41} & m_{42} & m_{43} &
m_{44}}\,\,\pmatrix{\delta ^{3} r \cr \delta ^{3} \theta \cr \delta ^{3}
\phi \cr \delta ^{3} t}=\epsilon ^{3}\,\pmatrix{D^r \cr D^\theta \cr D^\phi
\cr D^t}\,,  \label{matrixdelta3x}
\end{equation}
where the common factor $\epsilon ^{3}$ shows the explicit cubic dependence
of the third-order deviation $\delta ^{3}x^{\mu }$ on the first-order
deviation amplitude $n_{0}^{r}$ (or $n_{0}^{\theta }$). The functions $%
D^{r},D^{\theta }$, $D^{\phi }$ and $D^{t}$ are expressions depending on $M$%
, $R$, $a$, and $\sin $ and $\cos $ functions of $\omega s$, $\omega
_{\theta }s$, $3\omega $, $3\omega _{\theta }s$, $(\omega -2\omega _{\theta
})$, $(\omega +2\omega _{\theta })$, $(2\omega -\omega _{\theta })$ and $%
(2\omega +\omega _{\theta })$: 
\begin{equation}
D^{r}=D_{1}^{r}\cos (\omega s)+D_{3}^{r}\cos (3\omega s)+D_{-}^{r}\cos
[(\omega -2\omega _{\theta })s]+D_{+}^{r}\cos [(\omega +2\omega _{\theta
})s],  \label{Dr}
\end{equation}
\begin{equation}
D^{\theta }=D_{1}^{\theta }\cos (\omega _{\theta }s)+D_{3}^{\theta }\cos
(3\omega _{\theta }s)+D_{-}^{\theta }\cos [(2\omega -\omega _{\theta
})s]+D_{+}^{\theta }\cos [(2\omega +\omega _{\theta })s],  \label{Dtheta}
\end{equation}
\begin{equation}
D^{\phi }=D_{1}^{\phi }\cos (\omega s)+D_{3}^{\phi }\cos (3\omega
s)+D_{-}^{\phi }\cos [(\omega -2\omega _{\theta })s]+D_{+}^{\phi }\cos
[(\omega +2\omega _{\theta })s],  \label{Dphi}
\end{equation}
\begin{equation}
D^{t}=D_{1}^{t}\cos (\omega s)+D_{3}^{t}\cos (3\omega s)+D_{-}^{t}\cos
[(\omega -2\omega _{\theta })s]+D_{+}^{t}\cos [(\omega +2\omega _{\theta
})s].  \label{Dt}
\end{equation}

The functions $\cos \,(\omega \,s)$ and $\cos \,(\omega _{\theta }\,s)$
represent a new problem for the third-order deviation, as they are {\it %
resonance terms} whose angular frequency $\omega $ (or $\omega _{\theta }$)
is the same as the eigenvalue of the matrix-operator acting on the left-hand
side, yielding secular terms, proportional to $s$. To avoid unbounded
deviations, we can apply the Poincar\'{e}'s method \cite{Poincare} to take
into account possible perturbation of the basic frequency itself, replacing $%
\omega $ (or $\omega _{\theta }$) by an infinite series in powers of the
infinitesimal parameter, which in our case can be the ``eccentricity'' $%
\epsilon =\frac{{n_{0}^{r}}}{R}$: 
\begin{equation}
\omega \rightarrow \omega _{p}=\omega _{0}+\epsilon \,\omega _{1}+\epsilon
^{2}\,\omega _{2}+\epsilon ^{3}\,\omega _{3}+\dots \,,  \label{omega_p}
\end{equation}
where the new $\omega $ is renamed $\omega _{p}$ and $\omega _{0}$ is the
old $\omega $, and 
\begin{equation}
\omega _{\theta }\rightarrow \omega _{\theta p}=\omega _{\theta 0}+\epsilon
\,\omega _{\theta 1}+\epsilon ^{2}\,\omega _{\theta 2}+\epsilon ^{3}\,\omega
_{\theta 3}+\dots \,,  \label{omegatheta_p}
\end{equation}
where the new $\omega _{\theta }$ is renamed $\omega _{\theta p}$ and $%
\omega _{\theta 0}$ is the old $\omega _{\theta }$.

We shall build the complete differential equation for $x^{\mu }$, taking
together the harmonic oscillator equations for $\delta r$, $\delta ^{2}r$
and $\delta ^{3}r$: 
\[
\frac{d^{2}}{ds^{2}}(\delta r+\frac{\delta ^{2}r}{2}+\frac{\delta ^{3}r}{6}%
)+\omega _{0}^{2}(\delta r+\frac{\delta ^{2}r}{2}+\frac{\delta ^{3}r}{6})=%
\frac{\Delta ^{3}r_{0}}{6}+\frac{\Delta ^{3}r_{1}}{6}\cos (\omega _{p}s)+%
\frac{\Delta ^{2}r_{2}}{2}\cos (2\omega _{p}s) 
\]
\begin{equation}
+\frac{\Delta ^{3}r_{3}}{6}\cos (3\omega _{p}s)+\frac{\Delta ^{3}r_{-}}{6}%
\cos [(\omega _{p}-2\omega _{\theta p})s]+\frac{\Delta ^{3}r_{+}}{6}\cos
[(\omega _{p}+2\omega _{\theta p})s],
\end{equation}
and for $\delta \theta $, $\delta ^{2}\theta $ and $\delta ^{3}\theta $: 
\[
\frac{d^{2}}{ds^{2}}(\delta \theta +\frac{\delta ^{2}\theta }{2}+\frac{%
\delta ^{3}\theta }{6})+\omega _{\theta 0}^{2}(\delta \theta +\frac{\delta
^{2}\theta }{2}+\frac{\delta ^{3}\theta }{6})=\frac{\Delta ^{2}\theta _{-}}{2%
}\cos [(\omega _{p}-\omega _{\theta p})s]+\frac{\Delta ^{2}\theta _{+}}{2}%
\cos [(\omega _{p}+\omega _{\theta p})s] 
\]
\begin{equation}
+\frac{\Delta ^{3}\theta _{1}}{6}\cos (\omega _{\theta p}s)+\frac{\Delta
^{3}\theta _{3}}{6}\cos (3\omega _{\theta p}s)+\frac{\Delta ^{3}\theta _{-}}{%
6}\cos [(2\omega _{p}-\omega _{\theta p})s]+\frac{\Delta ^{3}\theta _{+}}{6}%
\cos [(2\omega _{p}+\omega _{\theta p})s].
\end{equation}
Then, developing both sides into a series of powers of the parameter $%
\epsilon $, we can not only recover the former differential equations for
the vectors $\delta x^{\mu },\,\delta ^{2}x^{\mu },\,\delta ^{3}x^{\mu }$,
but get also some algebraic relations defining the corrections $\omega
_{1},\,\omega _{2},$ $\omega _{\theta 1}$ and $\,\omega _{\theta 2}$, see
the Appendix 2. In the Schwarzschild limit, we have: 
\begin{equation}
\omega _{1}=0,\,\,\ \ \,\,\ \ \omega _{2}=-\frac{3}{4}\frac{M^{3/2}}{R^{5/2}}%
\frac{(6-37\frac{M}{R})}{\sqrt{1-\frac{3M}{R}}(1-\frac{6M}{R})^{3/2}}.
\label{omega12}
\end{equation}
so the new frequency corrected by the Poincar\'{e}'s method is simply: 
\begin{equation}
\omega _{p}=\frac{\sqrt{M}}{R^{3/2}}\frac{\sqrt{1-\frac{6M}{R}}}{\sqrt{1-%
\frac{3M}{R}}}-\frac{3\left( n_{0}^{r}\right) ^{2}M^{3/2}}{4R^{9/2}}\frac{%
(6-37\frac{M}{R})}{\sqrt{1-\frac{3M}{R}}(1-\frac{6M}{R})^{3/2}}.
\end{equation}

Finally, we can obtain that the first and second-order deviations are the
same, but with the new $\omega _{p}$ and $\omega _{\theta p}$; and the
third-order deviation $\delta ^{3}x^{\mu }$ is given by: 
\begin{equation}
\delta ^{3}r=\delta ^{3}r_{0}+\delta ^{3}r_{3}\cos (3\omega _{p}s)+\delta
^{3}r_{-}\cos [(\omega _{p}-2\omega _{\theta p})s]+\delta ^{3}r_{+}\cos
[(\omega _{p}+2\omega _{\theta p})s],  \label{d3r}
\end{equation}
\begin{equation}
\delta ^{3}\theta =\delta ^{3}\theta _{3}\cos (3\omega _{\theta p}s)+\delta
^{3}\theta _{-}\cos [(2\omega _{p}-\omega _{\theta p})s]+\delta ^{3}\theta
_{+}\cos [(2\omega _{p}+\omega _{\theta p})s],  \label{d3theta}
\end{equation}
\[
\delta ^{3}\phi =(\delta ^{3}\phi _{0})s+\delta ^{3}\phi _{1}\sin (\omega
_{p}s)+\delta ^{3}\phi _{3}\sin (3\omega _{p}s)+\delta ^{3}\phi _{-}\sin
[(\omega _{p}-2\omega _{\theta p})s] 
\]
\begin{equation}
+\delta ^{3}\phi _{+}\sin [(\omega _{p}+2\omega _{\theta p})s],
\label{d3phi}
\end{equation}
\[
\delta ^{3}t=(\delta ^{3}t_{0})s+\delta ^{3}t_{1}\sin (\omega _{p}s)+\delta
^{3}t_{3}\sin (3\omega _{p}s)+\delta ^{3}t_{-}\sin [(\omega _{p}-2\omega
_{\theta p})s] 
\]
\begin{equation}
+\delta ^{3}t_{+}\sin [(\omega _{p}+2\omega _{\theta p})s].  \label{d3t}
\end{equation}
We can choose the initial conditions of the differential solutions so that
the constants $\delta ^{3}r_{0}$, $\delta ^{3}\phi _{0}$ and $\delta
^{3}t_{0}$ are null. The Appendix 2 lists the explicit values of the above
coefficients.

The long expressions of the Kerr case are well simplified in the
Schwarzschild limit: 
\begin{equation}
\delta ^{3}\theta =0,\,\,\ \ \,\delta ^{3}r_{-}=\delta ^{3}r_{+}=\delta
^{3}\phi _{-}=\delta ^{3}\phi _{+}=\delta ^{3}t_{-}=\delta ^{3}t_{+}=0,
\end{equation}
and the non-null coefficients are: 
\begin{equation}
\delta ^{3}r_{3}=\frac{-9\left( n_{0}^{r}\right) ^{3}}{8R^{2}}\frac{(2-28%
\frac{M}{R}+97\frac{M^{2}}{R^{2}})}{(1-\frac{6M}{R})^{2}},
\end{equation}
\begin{equation}
\delta ^{3}\phi _{1}=\frac{9\left( n_{0}^{r}\right) ^{3}}{R^{3}}\frac{(1-%
\frac{7M}{R})}{(1-\frac{6M}{R})^{3/2}},\,\,\ \ \,\delta ^{3}\phi _{3}=\frac{%
\left( n_{0}^{r}\right) ^{3}}{4R^{3}}\frac{(26-336\frac{M}{R}+1083\frac{M^{2}%
}{R^{2}})}{(1-\frac{6M}{R})^{5/2}},
\end{equation}
\begin{equation}
\delta ^{3}t_{1}=\frac{3\left( n_{0}^{r}\right) ^{3}}{R^{2}}\sqrt{\frac{M}{R}%
}\frac{(2-19\frac{M}{R}+40\frac{M^{2}}{R^{2}}-36\frac{M^{3}}{R^{3}})}{(1-%
\frac{2M}{R})^{3}(1-\frac{6M}{R})^{3/2}},
\end{equation}
\begin{equation}
\delta ^{3}t_{3}=\frac{\left( n_{0}^{r}\right) ^{3}}{4R^{2}}\sqrt{\frac{M}{R}%
}\frac{(18-276\frac{M}{R}+1339\frac{M^{2}}{R^{2}}-2172\frac{M^{3}}{R^{3}}%
+1164\frac{M^{4}}{R^{4}})}{(1-\frac{2M}{R})^{3}(1-\frac{6M}{R})^{5/2}}.
\end{equation}
The same approach could be used in the second-order deviation calculations,
but it is not necessary because there are no resonances in the second-order
deviation equations.

Now we can compare the perihelion advance with the post-Newtonian limit, Eq.
(\ref{peradvance1}). The Schwarzschild limit gives the perihelion advance as 
\begin{eqnarray}
\Delta \varphi &=&\left( \frac{6\,\pi M}{R}+\frac{27\,\pi M^{2}}{R^{2}}+%
\frac{135\,\pi M^{3}}{R^{3}}+...\right) +\frac{\left( n_{0}^{r}\right) ^{2}}{%
R^{2}}\left( \frac{9\,\pi M}{R}+\frac{159\,\pi M^{2}}{2R^{2}}\right. 
\nonumber \\
&&\left. +\frac{585\,\pi M^{3}}{R^{3}}+...\right) +\frac{\left(
n_{0}^{r}\right) ^{4}}{R^{4}}\left( \frac{81\,\pi M^{2}}{2R^{2}}+\frac{%
594\,\pi M^{3}}{R^{3}}+...\right) +...
\end{eqnarray}
which agrees with the $\Delta \varphi $ of Eq. (\ref{peradvance1}) after
replacing the major semi-axis $a$ by the value of Eq. (\ref{semiaxisa}): 
\begin{equation}
\Delta \varphi =\frac{6\,\pi M}{R}+\frac{9\,\pi e^{2}M}{R}+...
\end{equation}
The perihelion advance of $\Delta \varphi $ in the Kerr case depends on $M$, 
$R$, $a$, $n_{0}^{r}$ and $n_{0}^{\theta }$, and can be described with high
accuracy by a long explicit expression, not shown here.

\section{Discussion}

We have further developed a new method for calculating geodesics in a
completely relativistic setting, using higher-order deviations without
introducing the Newtonian or post-Newtonian approximations.

The computation of first, second and third-order deviations for the Kerr
metric have shown that this method can be reduced to a straightforward
iteration of solving linear systems of differential equations with constant
coefficients.

The only complexity resides in the simplification of symbolic coefficients
of the deviations, which is successfully performed by means of symbolic
computing softwares \cite{Mathematica}-\cite{MathTensor}.

It is interesting to observe how at the very first level of approximation
the angular momentum density $a$ of the central body influences the
perihelion advance via two different effects, which are linear and quadratic
in $a$, respectively. The expressions linear in $a$ depend on the sign of
this parameter, i.e. on the relative sign of two rotations: that of the
central body, and the direction of the orbital motion -- a kind of
spin-orbital coupling. This is the so-called {\it dragging effect}
characteristic for General Relativity, which tends to raise the perihelion
advance if the rotation of the planet is in the same direction as the
rotation of the central body itself, and tends to decrease the perihelion
advance if these two rotations are opposite to each other. The terms
quadratic in $a$ represent an additional perihelion advance which is due to
the fact that the non-vanishing angular momentum of the central body is
perceived from the exterior as an extra energy, which by the equivalence
principle, may be considered as an extra mass $\delta M$ added to the
central mass $M$; therefore, it always tends to produce higher perihelion
advance.

It is worth stressing that these effects are absent in the first-order
post-Newtonian approximation. In this sense, our method gives a shorter way
enabling one to display certain effects, than the commonly used
post-Newtonian approach. Its convergence properties are very good, too, so
that there are physical situations when it is more appropriate. Consider a
small mass rotating quite close to a black hole, so that the quantity $%
GM/rc^{2}\simeq v^{2}/c^{2}$ is of the order of $0.1$; then the second
post-Newtonian effects are of the order $0.01$. Now, if the eccentricity of
the orbit is of the same order, i.e. $e=0.1$ then our third-order terms give
the precision of $0.001$, keeping the quasi exact functional dependence on
physical parameters $GM/r$ and $a$.

There are many possible applications and further developments. The
computation of fourth and higher-order deviations in Schwarzschild and Kerr
metrics can improve the accuracy for practical calculations, and is just a
matter of spending more time and computer resources because we have
developed a semi-automatic program for explicit calculation of higher-order
geodesic deviations.

The gravitomagnetic clock effect, i.e., the time-difference between the
orbits of two freely counter-revolving test particles around a central
rotating mass $M$, is an ideal target for high-order geodesic deviations, as
the usual approaches are limited to circular orbits \cite{Mashhoom1999} or
slowly rotating mass $M$ \cite{Mashhoom2001}, i.e., small values of $a$. So
the high-order geodesic deviations method has the potential to compute the
gravitomagnetic clock effect in the case of strong general relativistic
effects (large values of $M$ and $a$).

For practical applications involving the X-ray or gravitational radiation,
the dynamics of accretion disks, etc, it would be useful to generalize our
method to the case of initial orbits inclined w.r.t. the equatorial plane of
the rotating central mass $M$. For example, the Ref. \cite{Sibgatullin}
considers inclined orbits in the Kerr metric, with the constraint of
low-eccentricity orbits, i.e., up to the first-order deviation.

Still within the test particle concept, we can extend it for test bodies
carrying charge and/or internal spin, see Ref.~\cite{vanHolten2002} where
the first-order geodesic deviation is derived in the Reissner-Nordstr{\o }m
background field. We foresee that higher-order geodesic deviations for test
particles with spin can provide useful results to compare with the
experimental data of satellite gyroscopes.

We can also replace the background metric by some cylindrical or
axially-symmetric metric to investigate approximated models of star and
galaxy orbits, accretion disks, etc.

Almost all the work available in the domain of gravitational radiation is
based on post-Newtonian approximations \cite{Tanaka1}-\cite{Blanschaef}.
Within the higher-order geodesic deviations approach, one possibility is to
maintain the test particle mass $m$ negligible compared to $M$, and compute
the emission of gravitational radiation \cite{vanHolten1997}\ with some
formula better suited for Schwarzschild and Kerr metrics than the quadrupole
formula \cite{Peters}. Another possibility, more challenging, is to cope
with the finite-size of the mass $m$ by taking it into account with
appropriate perturbation of the background metric, then trying to repeat the
higher-order geodesic deviations calculations and finally employing a
modified gravitational radiation formula based on the perturbed metric.

\vskip0.4cm \noindent {\tbf Acknowledgments}\newline

R. C. Jr. would like to thank CAPES of Brazil for financial support.

\section*{Appendix 1}

The coefficients of the solution, in the Kerr metric case, for the
second-order geodesic deviation $\delta ^{2}x^{\mu }(s)$ are: 
\begin{equation}
\delta ^{2}r_{0}=0,\,\,\ \ \delta ^{2}r_{2}=-\frac{\left( n_{0}^{r}\right)
^{2}}{Rf_{6}}\left( 1-\frac{7M}{R}+\frac{10a\sqrt{M}}{R^{3/2}}-\frac{4a^{2}}{%
R^{2}}\right) ,
\end{equation}
\begin{equation}
\delta ^{2}\theta _{-}=-\delta ^{2}\theta _{+}=\frac{2n_{0}^{r}n_{0}^{\theta
}}{R}\frac{\sqrt{f_{3}}}{\sqrt{f_{6}}},
\end{equation}
\begin{eqnarray*}
\delta ^{2}\phi _{2r} &=&-\frac{2\left( n_{0}^{r}\right) ^{2}}{%
R^{2}f_{6}^{3/2}f_{2}^{2}}\left[ \left( 5-\frac{32M}{R}\right) \left( 1-%
\frac{2M}{R}\right) ^{2}-\frac{2a\sqrt{M}}{R^{3/2}}\left( 26-\frac{119M}{R}+%
\frac{126M^{2}}{R^{2}}\right) \right. \\
&&\left. +\frac{2a^{2}}{R^{2}}\left( 8-\frac{61M}{R}+\frac{66M^{2}}{R^{2}}%
\right) +\frac{2a^{3}\sqrt{M}}{R^{7/2}}\left( 5+\frac{21M}{R}\right) +\frac{%
a^{4}}{R^{4}}\left( 5-\frac{58M}{R}\right) +\frac{14a^{5}\sqrt{M}}{R^{11/2}}%
\right] ,
\end{eqnarray*}
\begin{equation}
\delta ^{2}\phi _{0}=0,\,\,\ \ \delta ^{2}\phi _{2\theta }=\frac{\left(
n_{0}^{\theta }\right) ^{2}}{\sqrt{f_{3}}}\left( 1-\frac{2a\sqrt{M}}{R^{3/2}}%
\right) ,
\end{equation}
\[
\delta ^{2}t_{0}=\frac{-3}{\sqrt{f_{1}}}\left[ \frac{\left( n_{0}^{r}\right)
^{2}}{R^{2}}\frac{\left( 1+\frac{M}{R}-\frac{4a\sqrt{M}}{R^{3/2}}+\frac{%
2a^{2}}{R^{2}}\right) }{2f_{2}}+\left( n_{0}^{\theta }\right) ^{2}\left( 
\frac{a\sqrt{M}}{R^{3/2}}-\frac{a^{2}}{R^{2}}\right) \right] , 
\]
\begin{eqnarray*}
&&\delta ^{2}t_{2r}=\frac{\left( n_{0}^{r}\right) ^{2}\sqrt{M}}{%
R^{3/2}f_{6}^{3/2}f_{2}^{2}}\left[ \left( 2-\frac{15M}{R}+\frac{14M^{2}}{%
R^{2}}\right) +\frac{a\sqrt{M}}{R^{3/2}}\left( 11+\frac{34M}{R}-\frac{64M^{2}%
}{R^{2}}\right) \right. \\
&&\left. -\frac{3a^{2}}{R^{2}}\left( 1+\frac{28M}{R}-\frac{42M^{2}}{R^{2}}%
\right) +\frac{2a^{3}\sqrt{M}}{R^{7/2}}\left( 28-\frac{33M}{R}\right) -\frac{%
3a^{4}}{R^{4}}\left( 4+\frac{7M}{R}\right) +\frac{29a^{5}\sqrt{M}}{R^{11/2}}-%
\frac{27a^{6}}{R^{6}}\right] ,
\end{eqnarray*}
\begin{equation}
\delta ^{2}t_{2\theta }=\left( n_{0}^{\theta }\right) ^{2}\frac{a^{2}\sqrt{M}%
}{R^{3/2}\sqrt{f_{3}}}.
\end{equation}

\section*{Appendix 2}

With the third-order deviation in the Kerr metric, we can determine the
values of: 
\begin{equation}
\omega _{1}=\omega _{\theta 1}=0,
\end{equation}
\begin{eqnarray}
\omega _{2} &=&-\frac{3\left( n_{0}^{\theta }\right) ^{2}a\sqrt{M}}{\left(
n_{0}^{r}\right) ^{2}\sqrt{R}}\frac{\left( \frac{\sqrt{M}}{\sqrt{R}}-\frac{a%
}{R}\right) f_{2}}{\sqrt{f_{1}}\sqrt{f_{6}}}-\frac{3\sqrt{M}}{4R^{3/2}}\frac{%
\left( \frac{\sqrt{M}}{\sqrt{R}}-\frac{a}{R}\right) ^{2}}{\sqrt{f_{1}}%
f_{2}f_{6}^{3/2}}\left[ \left( 1-\frac{2M}{R}\right) \left( 6-\frac{37M}{R}%
\right) \right.  \nonumber \\
&&\left. +\frac{2a\sqrt{M}}{R^{3/2}}\left( 27-\frac{62M}{R}\right) -\frac{%
a^{2}}{R^{2}}\left( 17-\frac{57M}{R}\right) +\frac{6a^{3}\sqrt{M}}{R^{7/2}}-%
\frac{7a^{4}}{R^{4}}\right] ,
\end{eqnarray}
\begin{equation}
\omega _{\theta 2}=-\frac{3a\sqrt{M}}{2R^{5/2}}\frac{\left( \frac{\sqrt{M}}{%
\sqrt{R}}-\frac{a}{R}\right) \left( 3-\frac{5M}{R}+\frac{2a^{2}}{R^{2}}%
\right) }{\sqrt{f_{1}}f_{2}\sqrt{f_{3}}}-\frac{3\left( n_{0}^{\theta
}\right) ^{2}a^{2}\sqrt{M}}{4\left( n_{0}^{r}\right) ^{2}R^{3/2}}\frac{%
\left( 1-\frac{4M}{R}+\frac{4a\sqrt{M}}{R^{3/2}}-\frac{a^{2}}{R^{2}}\right) 
}{\sqrt{f_{1}}\sqrt{f_{3}}},
\end{equation}
therefore the new frequencies corrected by the Poincar\'{e}'s method, which
are exact up to the second order w.r.t. the small parameter $\epsilon =\frac{%
{n_{0}^{r}}}{R}$, are given by: 
\begin{equation}
\omega _{p}=\omega _{0}+\frac{\left( n_{0}^{r}\right) ^{2}}{R^{2}}\,\omega
_{2}\,,\,\,\ \ \,\,\ \ \omega _{\theta p}=\omega _{\theta 0}+\frac{\left(
n_{0}^{r}\right) ^{2}}{R^{2}}\,\omega _{\theta 2}\,.
\end{equation}

The coefficients of the solution, in the Kerr metric case, for the
third-order geodesic deviation $\delta ^{3}x^{\mu }(s)$ are: 
\begin{eqnarray*}
\delta ^{3}r_{0}=0,\,\,\ \ \delta ^{3}r_{3} &=&\frac{-9\left(
n_{0}^{r}\right) ^{3}}{8R^{2}f_{6}^{2}}\left[ \left( 2-\frac{28M}{R}+\frac{%
97M^{2}}{R^{2}}\right) +\frac{4a\sqrt{M}}{R^{3/2}}\left( 10-\frac{69M}{R}%
\right) \right. \\
&&\left. -\frac{2a^{2}}{R^{2}}\left( 8-\frac{153M}{R}\right) -\frac{156a^{3}%
\sqrt{M}}{R^{7/2}}+\frac{31a^{4}}{R^{4}}\right] ,
\end{eqnarray*}
\begin{equation}
\delta ^{3}r_{-}=\frac{3n_{0}^{r}\left( n_{0}^{\theta }\right) ^{2}a^{2}}{%
4R^{2}}\frac{\sqrt{f_{6}}}{\sqrt{f_{3}}},\,\,\ \ \delta ^{3}r_{+}=\frac{%
3n_{0}^{r}\left( n_{0}^{\theta }\right) ^{2}a^{2}}{4R^{2}}\left( \frac{f_{6}+%
\sqrt{f_{3}}\sqrt{f_{6}}}{f_{3}+\sqrt{f_{3}}\sqrt{f_{6}}}\right) ,
\end{equation}
\[
\delta ^{3}\theta _{3}=\frac{-\left( n_{0}^{\theta }\right) ^{3}}{8f_{3}}%
\left[ 2-\frac{8a\sqrt{M}}{R^{3/2}}-\frac{3a^{2}}{R^{2}}\left( 1-\frac{4M}{R}%
\right) +\frac{12a^{3}\sqrt{M}}{R^{7/2}}-\frac{15a^{4}}{R^{4}}\right] , 
\]
\begin{equation}
\delta ^{3}\theta _{+}=-\delta ^{3}\theta _{-}=\frac{3\left(
n_{0}^{r}\right) ^{2}n_{0}^{\theta }}{4R^{2}}\left\{ \frac{4f_{3}}{f_{6}}-%
\frac{\sqrt{f_{3}}}{f_{6}^{3/2}}\left[ \left( 5-\frac{32M}{R}\right) +\frac{%
44a\sqrt{M}}{R^{3/2}}-\frac{17a^{2}}{R^{2}}\right] \right\} ,
\end{equation}
\begin{eqnarray*}
\delta ^{3}\phi _{1} &=&\frac{3\left( n_{0}^{r}\right) ^{3}}{%
R^{3}f_{6}^{3/2}f_{2}^{3}}\left[ 3\left( 1-\frac{2M}{R}\right) ^{3}\left( 1-%
\frac{7M}{R}\right) +\frac{a\sqrt{M}}{R^{3/2}}\left( 39-\frac{269M}{R}+\frac{%
580M^{2}}{R^{2}}-\frac{412M^{3}}{R^{3}}\right) \right. \\
&&-\frac{a^{2}}{R^{2}}\left( 14-\frac{169M}{R}+\frac{424M^{2}}{R^{2}}-\frac{%
332M^{3}}{R^{3}}\right) -\frac{a^{3}\sqrt{M}}{R^{7/2}}\left( 41-\frac{66M}{R}%
+\frac{12M^{2}}{R^{2}}\right) \\
&&\left. +\frac{a^{4}}{R^{4}}\left( 3+\frac{63M}{R}-\frac{134M^{2}}{R^{2}}%
\right) -\frac{a^{5}\sqrt{M}}{R^{11/2}}\left( 31-\frac{79M}{R}\right) +\frac{%
a^{6}}{R^{6}}\left( 4-\frac{17M}{R}\right) +\frac{a^{7}\sqrt{M}}{R^{15/2}}%
\right] ,
\end{eqnarray*}
\begin{eqnarray*}
\delta ^{3}\phi _{3} &=&\frac{\left( n_{0}^{r}\right) ^{3}}{%
4R^{3}f_{6}^{5/2}f_{2}^{3}}\left[ \left( 1-\frac{2M}{R}\right) ^{3}\left( 26-%
\frac{336M}{R}+\frac{1083M^{2}}{R^{2}}\right) +\frac{a\sqrt{M}}{R^{3/2}}%
\left( 518-\frac{6624M}{R}+\frac{27939M^{2}}{R^{2}}\right. \right. \\
&&\left. -\frac{48140M^{3}}{R^{3}}-\frac{29484M^{4}}{R^{4}}\right) -\frac{%
2a^{2}}{R^{2}}\left( 84-\frac{2479M}{R}+\frac{14109M^{2}}{R^{2}}-\frac{%
27456M^{3}}{R^{3}}+\frac{17604M^{4}}{R^{4}}\right) \\
&&-\frac{2a^{3}\sqrt{M}}{R^{7/2}}\left( 822-\frac{6251M}{R}+\frac{10737M^{2}%
}{R^{2}}-\frac{4422M^{3}}{R^{3}}\right) +\frac{3a^{4}}{R^{4}}\left( 69-\frac{%
410M}{R}-\frac{2971M^{2}}{R^{2}}\right. \\
&&\left. +\frac{5750M^{3}}{R^{3}}\right) -\frac{a^{5}\sqrt{M}}{R^{11/2}}%
\left( 777-\frac{11664M}{R}+\frac{15625M^{2}}{R^{2}}\right) +\frac{2a^{6}}{%
R^{6}}\left( 95-\frac{1965M}{R}+\frac{1176M^{2}}{R^{2}}\right) \\
&&\left. +\frac{6a^{7}\sqrt{M}}{R^{15/2}}\left( 55+\frac{463M}{R}\right) +%
\frac{a^{8}}{R^{8}}\left( 45-\frac{1474M}{R}\right) +\frac{225a^{9}\sqrt{M}}{%
R^{19/2}}\right] ,
\end{eqnarray*}
\[
\delta ^{3}\phi _{0}=0,\,\,\ \ \delta ^{3}\phi _{-}=\frac{3n_{0}^{r}\left(
n_{0}^{\theta }\right) ^{2}}{2R}\left[ \frac{\left( 2-\frac{4a\sqrt{M}}{%
R^{3/2}}\right) }{\sqrt{f_{6}}}+\frac{a^{2}f_{4}}{R^{2}f_{2}\sqrt{f_{3}}}%
\right] , 
\]
\begin{equation}
\delta ^{3}\phi _{+}=\frac{3n_{0}^{r}\left( n_{0}^{\theta }\right) ^{2}}{2R}%
\left[ \frac{\left( 2-\frac{4a\sqrt{M}}{R^{3/2}}\right) }{\sqrt{f_{6}}}-%
\frac{a^{2}f_{4}}{R^{2}f_{2}\sqrt{f_{3}}}\right] ,
\end{equation}
\begin{eqnarray*}
&&\delta ^{3}t_{1}=\frac{3\left( n_{0}^{r}\right) ^{3}\sqrt{M}}{%
R^{5/2}f_{6}^{3/2}f_{2}^{3}}\left[ \left( 2-\frac{19M}{R}+\frac{40M^{2}}{%
R^{2}}-\frac{36M^{3}}{R^{3}}\right) +\frac{a\sqrt{M}}{R^{3/2}}\left( 13+%
\frac{26M}{R}-\frac{140M^{2}}{R^{2}}+\frac{168M^{3}}{R^{3}}\right) \right. \\
&&-\frac{a^{2}}{R^{2}}\left( 5+\frac{103M}{R}-\frac{396M^{2}}{R^{2}}+\frac{%
412M^{3}}{R^{3}}\right) +\frac{a^{3}\sqrt{M}}{R^{7/2}}\left( 73-\frac{332M}{R%
}+\frac{332M^{2}}{R^{2}}\right) -\frac{a^{4}}{R^{4}}\left( 15-\frac{91M}{R}%
\right. \\
&&\left. \left. +\frac{12M^{2}}{R^{2}}\right) +\frac{a^{5}\sqrt{M}}{R^{11/2}}%
\left( 11-\frac{134M}{R}\right) -\frac{a^{6}}{R^{6}}\left( 7-\frac{79M}{R}%
\right) -\frac{17a^{7}\sqrt{M}}{R^{15/2}}+\frac{a^{8}}{R^{8}}\right] ,
\end{eqnarray*}
\begin{eqnarray*}
&&\delta ^{3}t_{3}=\frac{\left( n_{0}^{r}\right) ^{3}\sqrt{M}}{%
4R^{3}f_{6}^{5/2}f_{2}^{3}}\left[ \left( 18-\frac{276M}{R}+\frac{1339M^{2}}{%
R^{2}}-\frac{2172M^{3}}{R^{3}}+\frac{1164M^{4}}{R^{4}}\right) +\frac{2a\sqrt{%
M}}{R^{3/2}}\left( 120-\frac{762M}{R}\right. \right. \\
&&\left. -\frac{755M^{2}}{R^{2}}+\frac{5124M^{3}}{R^{3}}-\frac{4332M^{4}}{%
R^{4}}\right) -\frac{a^{2}}{R^{2}}\left( 74-\frac{210M}{R}-\frac{13185M^{2}}{%
R^{2}}+\frac{38736M^{3}}{R^{3}}-\frac{29484M^{4}}{R^{4}}\right) \\
&&+\frac{4a^{3}\sqrt{M}}{R^{7/2}}\left( 85-\frac{4404M}{R}+\frac{12069M^{2}}{%
R^{2}}-\frac{8802M^{3}}{R^{3}}\right) -\frac{a^{4}}{R^{4}}\left( 105-\frac{%
10426M}{R}+\frac{21699M^{2}}{R^{2}}\right. \\
&&\left. -\frac{8844M^{3}}{R^{3}}\right) -\frac{2a^{5}\sqrt{M}}{R^{11/2}}%
\left( 1461+\frac{2902M}{R}-\frac{8625M^{2}}{R^{2}}\right) +\frac{a^{6}}{%
R^{6}}\left( 309+\frac{10158M}{R}-\frac{15625M^{2}}{R^{2}}\right) \\
&&\left. -\frac{48a^{7}\sqrt{M}}{R^{15/2}}\left( 83-\frac{49M}{R}\right) +%
\frac{a^{8}}{R^{8}}\left( 547+\frac{2778M}{R}\right) -\frac{1474a^{9}\sqrt{M}%
}{R^{19/2}}+\frac{225a^{10}}{R^{10}}\right] ,
\end{eqnarray*}
\[
\delta ^{3}t_{0}=0,\,\,\ \ \delta ^{3}t_{-}=\frac{3n_{0}^{r}\left(
n_{0}^{\theta }\right) ^{2}a^{2}\sqrt{M}}{2R^{5/2}}\left[ \frac{-4}{\sqrt{%
f_{6}}}+\frac{f_{5}}{f_{2}\sqrt{f_{3}}}\right] , 
\]
\begin{equation}
\delta ^{3}t_{+}=\frac{3n_{0}^{r}\left( n_{0}^{\theta }\right) ^{2}a^{2}%
\sqrt{M}}{2R^{5/2}}\left[ \frac{-4}{\sqrt{f_{6}}}-\frac{f_{5}}{f_{2}\sqrt{%
f_{3}}}\right] .
\end{equation}

\end{document}